\documentclass[graybox]{svmult}

\usepackage{type1cm}        % activate if the above 3 fonts are
\usepackage{makeidx}         % allows index generation
\usepackage{graphicx}        % standard LaTeX graphics tool
\usepackage{multicol}        % used for the two-column index
\usepackage[bottom]{footmisc}% places footnotes at page bottom

\usepackage{newtxtext}       % 
\usepackage{newtxmath}       % selects Times Roman as basic font

\makeindex             % used for the subject index
\begin{document}

\title*{Observational Constraints on the Common Envelope Phase}
%\titlerunning{Common Envelope Phase}
\author{David Jones}
\institute{David Jones \at Instituto de Astrof\'isica de Canarias, E-38205 La Laguna, Tenerife, Spain\\
Departamento de Astrof\'isica, Universidad de La Laguna, E-38206 La Laguna, Tenerife, Spain 
\email{djones@iac.es}
}
\maketitle

\abstract{The common envelope phase was first proposed more than forty years ago to explain the origins of evolved, close binaries like cataclysmic variables.  It is now believed that the phase plays a critical role in the formation of a wide variety of other phenomena ranging from type \textsc{i}a supernovae through to binary black holes, while common envelope mergers are likely responsible for a range of enigmatic transients and supernova imposters.  Yet, despite its clear importance, the common envelope phase is still rather poorly understood.  Here, we outline some of the basic principles involved, the remaining questions as well as some of the recent observational hints from common envelope phenomena - namely planetary nebulae and luminous red novae - which may lead to answering these open questions.}

\section{Preface}
\label{sec:pre}

It is important to highlight from the outset that the common envelope (CE) and its progeny have been the subject of constant and vibrant study, both observational and theoretical, for many years.  As a result, several excellent reviews have already been written \cite{iben93,taam00,webbink08,ivanova13}.  The review presented here is intended to be complementary to these, with an emphasis on the (sometimes puzzling) observations which may hold the key to understanding the CE.

\section{Introduction}
\label{sec:intro}

The Roche model (named for the nineteenth century French astronomer \'Edouard Roche) describes the gravitational potential of a close-binary system assuming that the two stars are well represented by synchronously-rotating point masses in a circular orbit \cite{kopal59}.  While rather simplistic, this model is still a relatively good representation of many close binaries with detailed numerical simulations generally required to complement and verify the model \cite{dermine09,deschamps13}.

\begin{figure}[t]
\includegraphics[width=\textwidth]{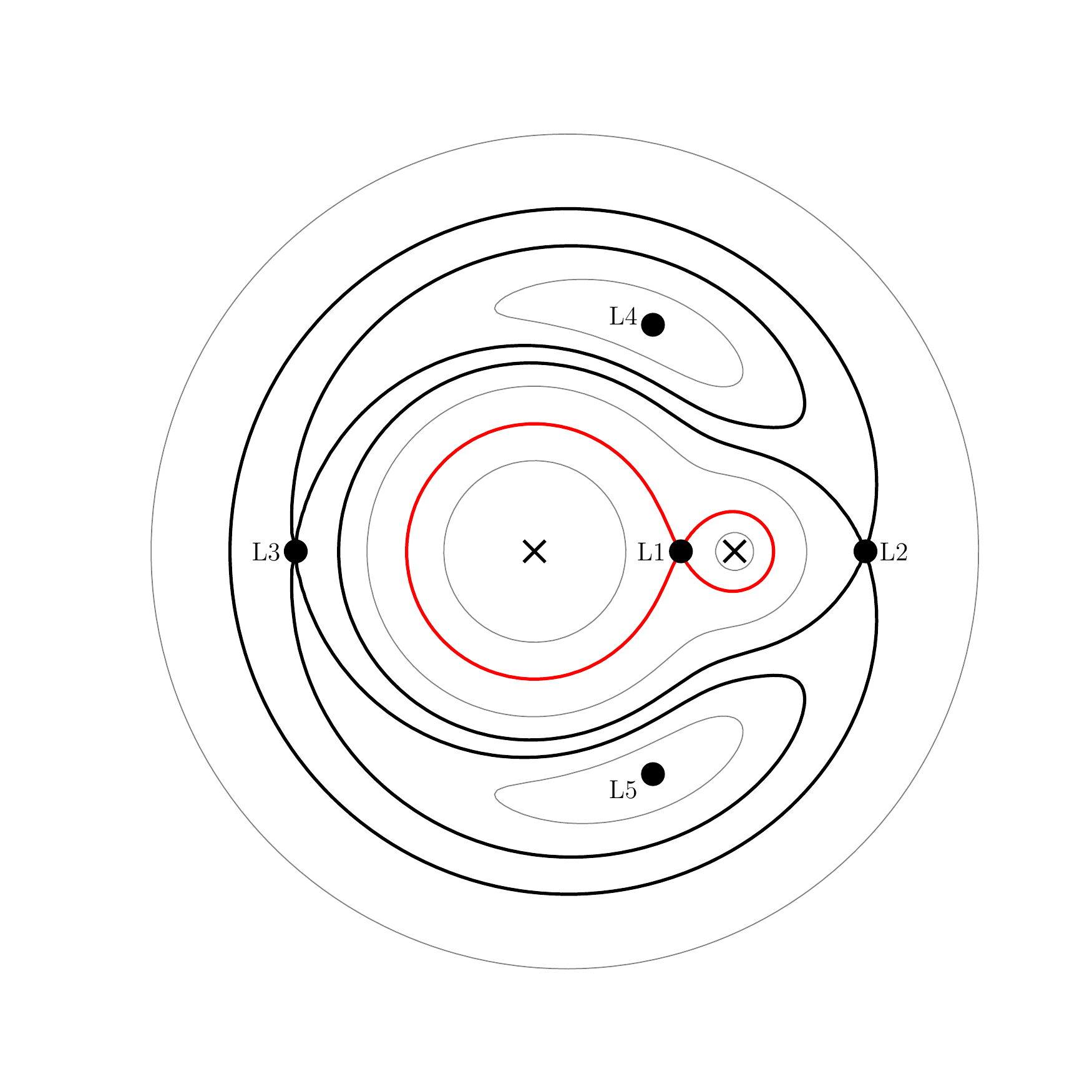}
\caption{The lines of equipotential, including the Roche lobe (marked in red), around a binary system with a mass ratio of 5 (i.e.\ the left-most star of the binary is five times as massive as its companion).  The Langrangian points are also marked (L1 through L5) as are the centres of mass of the two stars (black crosses).}
\label{fig:roche}       % Give a unique label
\end{figure}

The Roche model predicts the existence of five local minima of gravitational potential surrounding the binary, known as the Lagrangian points (named for Joseph-Louis Lagrange who discovered the fourth and fifth points shortly after Leonhard Euler discovered the first three).  Furthermore, the Roche model also calculates the existence of an equipotential surface enclosing each star which represents the largest extent at which a point mass could be gravitationally bound to that star (rather than to the binary system as a whole or unbound completely).  These so-called Roche lobes meet at the first Lagrangian point.  The configuration of the Roche lobes and five Lagrangian points is highlighted in Fig.~\ref{fig:roche}.  The extent of the Roche lobe is a function of the orbital separation and mass ratio of the binary components (being larger for the more massive component).  A commonly used approximation for the Roche lobe radius, $R_{1,RL}$, is
\begin{eqnarray}
\label{eqn:eggleton}
    R_{1,RL}&=&a \; r_{1,RL}\\
    r_{1,RL} &\approx& \frac{0.49q^{2/3}}{0.6q^{2/3}+\mathrm{ln}(1+q^{1/3})}
\end{eqnarray}
where
\begin{equation}
    q=\frac{M_1}{M_2}
\end{equation}
and $a$ is the separation of the stellar centres of mass, $M_1$ is the mass of the star for which we are calculating the Roche lobe radius and $M_2$ is the mass of its companion \cite{eggleton83}.  This Roche lobe radius is conceptually important in understanding the CE as it represents the critical stellar radius, beyond which the star will begin to transfer material on to its companion via L$_1$.  A process known as Roche-lobe overflow (RLOF).  This can occur at various stages during the evolution of the star depending on the component masses as well as the binary orbital period (more massive stars at shorter orbital periods are more likely to fill their Roche lobes earlier during their evolution).

If RLOF occurs while the donor is on the main sequence, this is generally referred to as Case A RLOF.  If the star is on its first ascent of the giant branch (i.e. a red giant), then it would be Case B, and while on the second ascent (asymptotic giant branch) it is Case C.  Given the change in stellar radius between these evolutionary stages, Case A occurs at much shorter orbital periods than Case B and Case C (see Sec.~ \ref{sec:conditions} for further discussion).  These are important definitions as the evolutionary phase of the donor can have an important impact on its reaction to the mass loss.  For example, a star with a large, radiative envelope is likely to shrink in response to mass loss (and thus, perhaps, recede away from filling its Roche lobe, ending the mass transfer), while those with deep convective envelopes are more likely to expand \cite{eggleton00}.  In this case, a positive feedback loop is initiated with the star continually expanding in response to mass loss, leading to yet more mass loss.  The response of the system as a whole then depends on the reaction of the accretor.  If the companion can accept and thermally-adjust to this accreted material (i.e. if the mass transfer is sufficiently slow \cite{prialnik85}), then the mass transfer could potentially be stable. Otherwise, the accretor will quickly be driven out of thermal equilibrium and expand to fill its Roche lobe, with any further mass lost from the donor's envelope forming a common envelope of material surrounding the binary.  The response of the companion is actually rather less dependent on its own properties or evolutionary state than it is on the mass transfer rate.  As such, the key ingredient in forming a CE (or not) is the response of the mass-losing donor star - which must act to continue to overflowing its Roche lobe.

\begin{figure}[t]
\sidecaption[t]
\includegraphics[width=6.7cm]{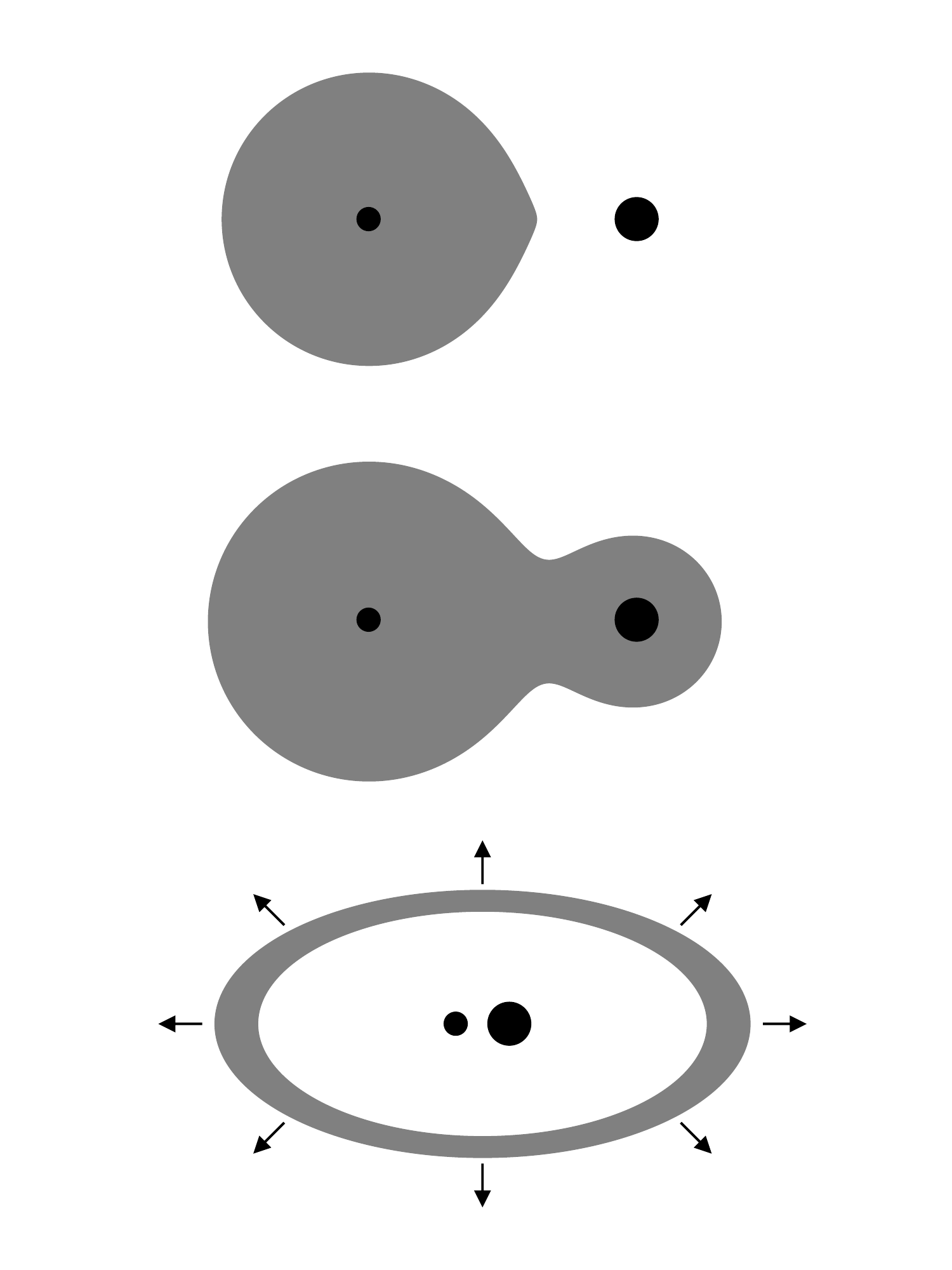}
\caption{A toy model of the common envelope.  The phase begins when the primary (left) begins to overflow its Roche lobe transferring material to its companion (right), as shown by the configuration in the top panel.  If neither the primary nor the secondary can adjust to the mass transfer, the secondary will also fill its Roche lobe resulting in the formation of a common envelope of material surrounding the secondary and the primary's core (middle panel).  Drag forces then transfer orbital energy and angular momentum from the binary to the envelope leading to its ejection while reducing the binary orbital separation (bottom panel).}
\label{fig:CE}       % Give a unique label
\end{figure}

Once inside the CE, the envelope of the donor star is almost certainly not co-rotating with the orbital motion of the companion \cite{paczynski76} -- even if tidal interactions had driven the pre-CE system into corotation, instabilities such as the Darwin instability will lead to at least some non-corotation \cite{darwin79}.  Drag forces between the orbiting companion and the surrounding CE then cause the companion to spiral in towards the core of the primary, transferring orbital energy and orbital angular momentum to the envelope.  The end result of this process being the dramatic reduction of the orbital period (perhaps even to merger) and ejection of the envelope (or some fraction of it in the case of a merger).  A cartoon of the main steps in this process is shown in Fig.~\ref{fig:CE}.  

As pointed out by \cite{paczynski76}, the CE provides a clear evolutionary pathway towards the formation of close binaries with an evolved component which would have been too large while on the giant branch to exist in the current orbital configuration - highlighting the example of V471 Tau which comprises a K-type main sequence star in a 12.5-hr orbit with a 0.8~M$_\odot$ white dwarf (WD).  A system which would have necessitated a $\sim$10 year orbit in order to accommodate the full asymptotic giant branch radius of the WD progenitor.

In discussing the CE hypothesis, \cite{paczynski76} concluded that following its ejection:
\begin{quotation}
We are left with two small stars accreting whatever hydrogen rich matter is left within their Roche lobes. At this time the degenerate core with the remaining envelope has a structure which is identical to that of a nucleus of a planetary nebula.  This hot star  will ionize the expanding envelope. As a result we should see a planetary nebula with a close binary as its nucleus.
\end{quotation}
Indeed, the discovery of the first close-binary planetary nebula (PN) nucleus, which they considered ``important support for the evolutionary scenario'' of the CE, was found later that year \cite{bond76}.  We will return to the importance of PNe in understanding the CE later, but first we must cover more of the theory and mathematical prescriptions used to study the CE.

\section{Conditions for a common envelope}
\label{sec:conditions}

For a binary to experience a CE, clearly, one component must initiate the process by filling its Roche lobe - this can occur for a number of reasons.  Dynamical interactions, for example with a third body leading to Kozai-Lidov interactions \cite{shappee13,stephan19}, can shrink the orbital separation sufficiently that one star becomes Roche lobe filling.  In the majority of cases, however, the CE will be initiated due to the radius evolution of the star.  This gives us important clues as to when a given star could fill its Roche lobe and initiate the CE, for example, as the radius evolution on the main sequence (MS) is minimal one would expect the vast majority of CE's to occur when the primary has evolved off the MS.  Similarly, the much smaller MS radii would also imply far smaller orbital separations, again restricting the likelihood of a MS CE event.

\begin{figure}[b]
\sidecaption[t]
\includegraphics[width=6.7cm]{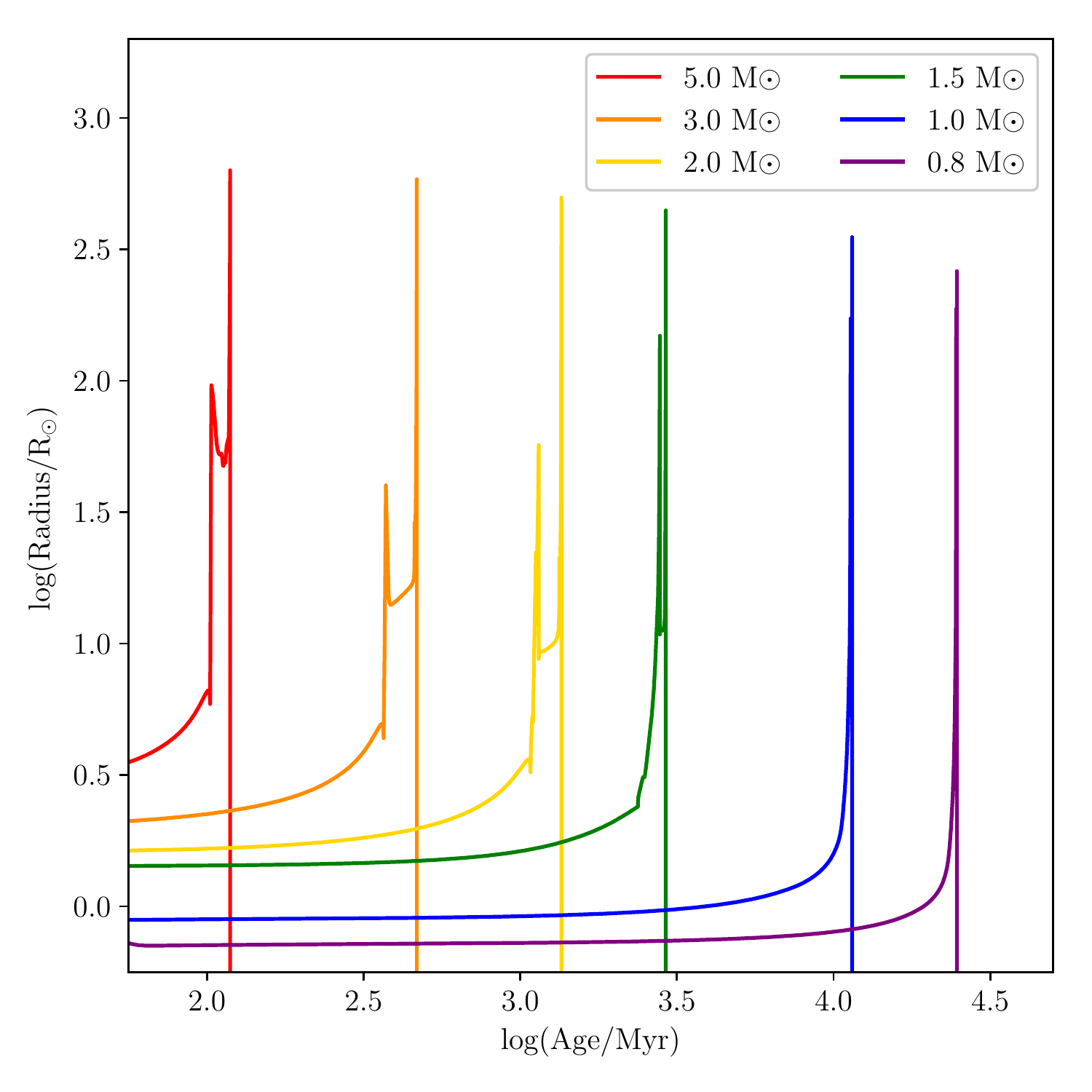}
\caption{The evolution of stellar radius as a function of time for a variety of initial stellar masses, (based on MIST tracks \cite{paxton11,choi16}).  Note that while the maximum radius (always at the tip of the AGB) increases as a function of initial mass, lower mass stars tend to achieve larger radii while on the RGB greatly increasing the likelihood of experiencing a CE while on the RGB.}
\label{fig:rad}       % Give a unique label
\end{figure}

Looking at the evolution of stellar radii (see e.g. Fig.~\ref{fig:rad}), one can see that more massive stars reach larger maximum radii at the tip of the AGB than their lower mass counterparts - thus increasing the range of orbital periods for which such a star will fill its Roche lobe and thus the likelihood of experiencing a CE event \footnote{This is roughly assuming that the orbital period distribution is not strongly dependent on the primary mass, which may not be the case \cite{moe17}.}.  Perhaps more interestingly, lower mass stars tend to reach larger radii while on the RGB, with their maximum RGB radii being rather comparable to their maximum radii while on the AGB.  This means that for lower mass stars, the vast majority of CE events will occur while on the RGB while the envelope itself is more massive and more bound - likely impacting on the energetics of the CE and the likelihood of ejection/binary survival (see Sec.~\ref{sec:alpha}).  More massive stars, however reach significantly larger radii while on the AGB compared to the RGB, thus these stars are more likely to experience AGB CE events particularly given the observed orbital period distribution \cite{moe17}.

\begin{figure}[t]
\sidecaption[t]
\includegraphics[width=6.7cm]{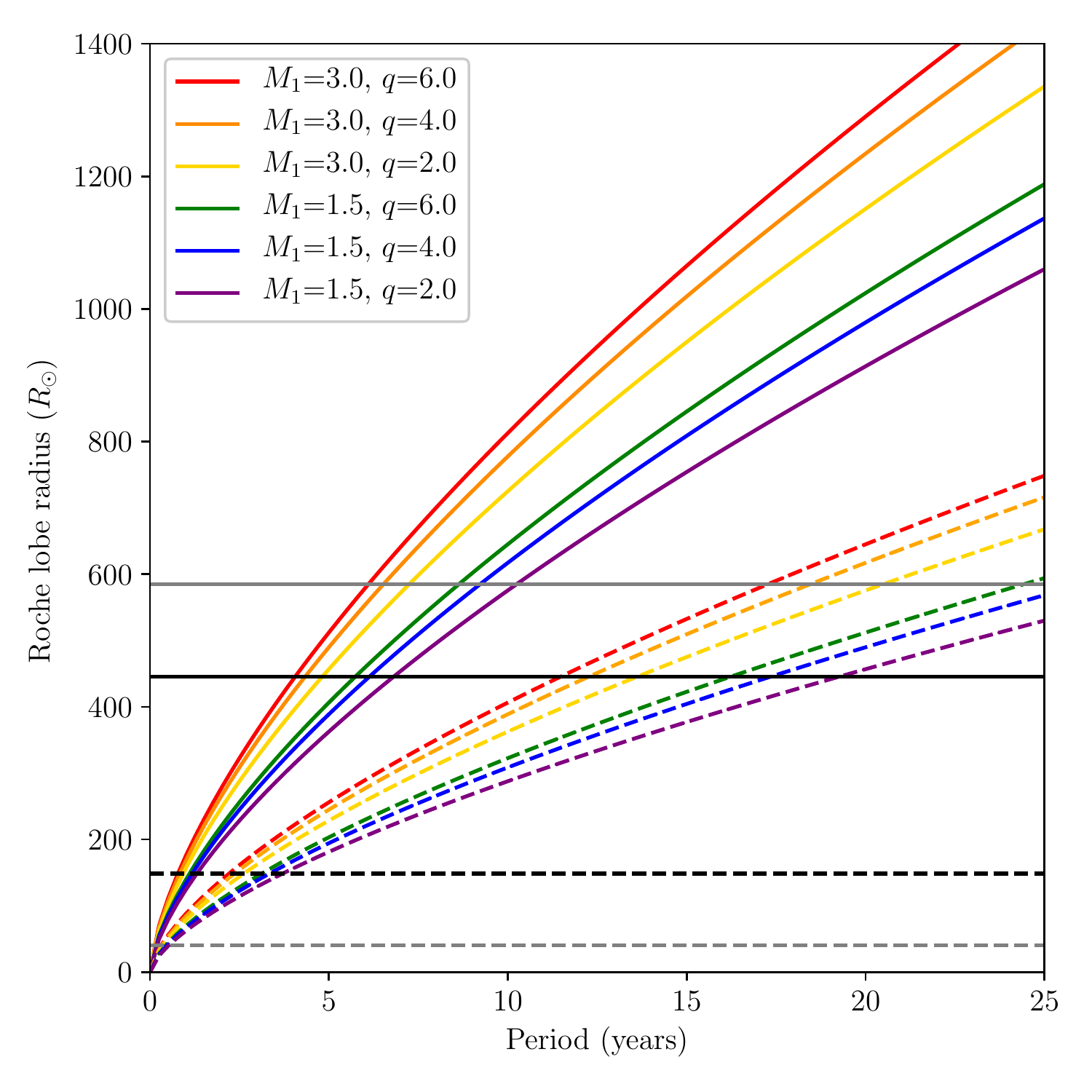}
\caption{The Roche lobe radius as a function of orbital period for various combinations of primary mass ($M_1$) and mass ratio ($q$).  The solid lines show configurations with zero eccentricity while the dashed lines are for eccentricity, $e=0.5$.  The horizontal lines represent the maximum RGB (dashed) and AGB (solid) radii for the primary masses considered, with the gray lines being $M_1=3.0 M_\odot$ while the black lines show $M_1=1.5 M_\odot$.}
\label{fig:rocherad}       % Give a unique label
\end{figure}

Somewhat obviously, the likelihood of entering a CE is not only a function of the primary's radius, but also on the parameters of the binary - namely the orbital separation and mass ratio (as shown in Eqn.~\ref{eqn:eggleton}).  To highlight these dependencies, in Fig.~\ref{fig:rocherad}, the Roche lobe radius is plotted as a function of orbital period for a range of binary configurations.  It is clear that all masses of primary will be Roche lobe filling at longer orbital periods for smaller mass ratios.  In terms of primary mass, more massive primaries have larger Roche lobe radii but also reach larger maximum radii on the AGB counteracting the effect - in the systems considered this leads to a similar spread of maximum orbital periods for which the systems will experience Roche lobe overflow, however this is not a general rule. It is important to also consider the possible influence of orbital eccentricity, which acts to reduce the orbital separation at periastron passage (by a factor $1-e$) and increase the likelihood of Roche lobe overflow.  Note, however, that as the system evolves tidal dissipation should act to reduce orbital eccentricity unless some eccentricity pumping mechanism is at work in the system \cite{vos15,saladino19}.  

Thus far, we have considered only the conditions required for a star to fill its Roche lobe but this is not the only necessity for a CE.  The resulting mass transfer must also be dynamically unstable - a runaway process where the reaction of the primary to mass loss is to continue to overflow its Roche lobe.  If the mass transfer is conservative, the primary's Roche-lobe radius evolves only with the mass ratio and thus the stability of mass transfer only depends on the mass ratio and how the primary's radius reacts to the mass loss.  Often, the primary's radius is considered to be proportional to some exponent of its mass, $R_d \propto M_1^{\zeta_*}$, with the Roche lobe radius behaving similarly but with a different exponent, $R_L \propto M_1^{\zeta_{RL}}$.  The mass transfer would thus be unstable if $\zeta_{RL} > \zeta_*$.

Given that the Roche lobe radius is principally dependent on the mass ratio but also on the total system mass and orbital period, the value of $\zeta_{RL}$ depends heavily on whether the mass transfer is conservative.  In the fully conservative case, $\zeta_{RL}$ is proportional to the mass ratio and often approximated by \cite{eggleton00}:
\begin{equation}
    \zeta_{RL} \approx 2.13q - 1.67.
\end{equation}
As such, a condition for unstable mass transfer becomes:
\begin{equation}
    q > \frac{\zeta_*}{2.13} + 0.79.
\end{equation}
The non-conservative case is rather more complicated but, in general, leads to lower values of $\zeta_*$ making stable mass transfer more likely.

The response of the donor's radius, characterised by $\zeta_*$, is more complicated than that of its Roche lobe and depends on the mass transfer time scale. For fully convective stars, like low-mass main sequence stars, one can approximate the star as a polytrope of index 1.5, leading to $\zeta_*=-1/3$.  Therefore, such stars will always respond to mass loss by increasing in size, thereby guaranteeing unstable mass transfer.  However, RGB and AGB stars are more complicated with convective envelopes which necessitate the use of condensed polytropes or, more properly, complete stellar models.  For these stars, $\zeta_*$ is always larger and quite frequently positive.  An approximation often used for red giants is:
\begin{equation}
\label{eqn:zetastar}
    3\zeta_* = 2 \frac{q_c}{1-q_c} - \frac{1-q_c}{1+2q_c},
\end{equation}
where $q_c$ is the fraction of stellar mass in the giant donor's core \cite{soberman97}.  However, this approximation assumes hydrostatic equilibrium - an assumption which only holds if the dynamical timescale is much shorter than the mass transfer timescale (which is almost certainly not the case).  This dramatically alters the stability condition for the mass transfer.  Binary stellar evolution modelling by \cite{woods11} showed that for a 5~M$_\odot$ giant donor with a 0.86~M$_\odot$ core, the mass transfer was found to be stable for mass ratios up to q=1.47, while using Eqn.~\ref{eqn:zetastar} one would predict unstable mass transfer for $q>0.75$.  As such, strong conclusions with regards mass transfer stability cannot be drawn except for specific cases which have been modelled using detailed stellar evolutionary codes.  In any case, it seems likely that only rather large mass ratios, $q\gtrsim2$, could lead to a CE \cite{passy12b,pavlovskii15}.

\section{Common envelope energetics}
\label{sec:alpha}

The outcome of the CE -- be that merger or a surviving short-period binary -- depends on whether the energy transferred to the envelope was sufficient to unbind it.  This permits us to define a parameter, known as the common envelope efficiency, which relates the change in orbital energy due as a result of the CE to the binding energy of the envelope \cite{tutukov79,webbink84}.  One can write the change in orbital energy (i.e. the sum of both gravitational potential and kinetic energies of both bodies) as

\begin{equation}
\label{eqn:e_orb1}
    \Delta E_\mathrm{orb} = G\; \bigg( \frac{M_{1,f} M_{2,f}}{2a_f} - \frac{M_{1,i} M_{2,i}}{2a_i} \bigg)
\end{equation}
where the subscripts 1 and 2 refer to the Roche-lobe-filling star (primary/donor) and its companion (secondary/accretor), respectively, while $i$ and $f$ denote initial (pre-CE) and final (post-CE) values. The prescription used here defines the initial orbital energy as being that between the entirety of the overflowing star and its companion at the initial separation \cite{webbink84}. Alternative prescriptions define it as that of the primary's core and the secondary (i.e. the $M_{1,i}$ term in equation \ref{eqn:e_orb1} becomes the primary's core mass, $M_c$, only), essentially assuming that the envelope has already engulfed the system and is now bound to the secondary and the primary's core  rather than to the primary only \cite{iben93,yungelson94}.  This definition is sometimes referred to as the Iben-Livio-Yungelson formulation \cite{zorotovic10} and, as we will shortly see, also implies an alternative prescription for the envelopes binding energy. 

Assuming that the companion does not grow in mass during the CE (i.e.\ $M_{2,f}=M_{2,i}=M_2$), and that the all of the overflowing star's envelope ($M_e$) is ejected, leaving behind only its core ($M_c$), the change in orbital energy can be rewritten
\begin{equation}
\label{eqn:e_orb2}
    \Delta E_\mathrm{orb}= G\; \bigg( \frac{M_{c} M_{2}}{2a_f} - \frac{(M_{c}+M_{e}) M_{2}}{2a_i} \bigg).
\end{equation}

Some fraction, $\alpha$, of this liberated orbital energy is transferred to the envelope.  Thus, assuming that there are no other potential sources of energy (an assumption which likely does not hold and which we will discuss in subsection \ref{sec:alpha_changers}), this fraction ($\alpha\Delta E_\mathrm{orb}$) should at least be equal to the binding energy of the envelope ($E_b$) in order for the binary to exit the CE without merging,
\begin{equation}
\label{eqn:alpha}
    E_b = \alpha \Delta E_\mathrm{orb}.
\end{equation}

As mentioned before, the definition of the binding energy is dependent on the formalism employed.  For the aforementioned Iben-Livio-Yungelson formulation, the binding energy is taken to be the gravitational energy between the primary's envelope and the combined mass of the primary's core and the companion.  The more-commonly-used alternative is the Podsialowski-Rappaport-Han formulation, which is more consistent with the definition of the change of orbital energy in equations \ref{eqn:e_orb1} and \ref{eqn:e_orb2}.  Under this formulation, the envelope is considered to be bound only to the primary,  and is often approximated as the gravitational energy between the envelope and primary mass,

\begin{equation}
\label{eqn:eb_m1}
E_b = G \frac{M_1 M_e}{\lambda r_{1,RL}}
\end{equation}
where $\lambda$ is of order unity and describes the radial mass distribution of the primary's envelope.  Here, the radius is assumed to be the Roche lobe radius as the star must be overflowing at the start of the CE.  More formally, rather than estimating the binding energy of the envelope, one can integrate envelope mass from stellar structure.  Using such models, the following approximation can be derived \cite{demarco11},
\begin{equation}
\label{eqn:eb_int}
    E_b=G \frac{(\frac{M_e}{2}+M_c)M_e}{\lambda r_{1,RL}},
\end{equation}
similar but subtly different to the more simplistic version shown in equation \ref{eqn:eb_m1}.

Combining equations \ref{eqn:e_orb2}, \ref{eqn:alpha} and \ref{eqn:eb_int} gives:
\begin{equation}
\label{eqn:alphafull}
    \frac{\big(\frac{M_e}{2}+M_c\big)M_e}{\lambda r_{1,RL}} = \alpha \bigg( \frac{M_{c} M_{2}}{2a_f} - \frac{(M_{c}+M_{e}) M_{2}}{2a_i} \bigg)
\end{equation}
in which $M_c$ (the core mass of the primary, assumed to also be the post-CE remnant mass), $M_2$ (the secondary mass, assumed to be unchanged by the CE) and $a_f$ (the post-CE orbital separation\footnote{The assumption that the current separation is equal to the immediately post-CE separation only holds for ``young'' systems, which have experienced negligible angular momentum loss since the CE.  For older systems, one must account for the influence of disrupted magnetic braking on the observed orbital period \cite{davis12}.}) are all observables (or can at least be derived from observations by, for example, combined light and radial velocity curve modelling).  Furthermore, using stellar evolutionary models, one can find possible progenitors for the primary which would present with consistent core masses and pre-CE radii, constraining $M_e$ ($\equiv M_1 - M_c$), $r_{1,RL}$ and $a_f$ \cite{zorotovic10,demarco11,davis12}.  The only remaining variable needed in order to derive the CE efficiency for a given system is the stellar structure parameter, $\lambda$, which is frequently assumed to be a constant (e.g. 0.5 \cite{dekool87}) or simply left incorporated in the efficiency as the product $\alpha \lambda$ \cite{nelemans05,zorotovic10}.  Alternatively, the parameter can be estimated based on stellar evolutionary models \cite{dewi00,demarco11}.  Ultimately, this permits a ``reconstruction'' of the CE phase and the estimation of the efficiency, $\alpha$, for individual systems (as opposed to using a population synthesis approach \cite{toonen13}).

\subsection{Other factors that may impact $\alpha$}
\label{sec:alpha_changers}

In the previous section, we have assumed that only orbital energy is responsible for the unbinding of the CE (as has traditionally been considered).  However, there are likely many other factors which may play a role in helping to remove the envelope.  The most obvious of these is the thermal energy of the envelope itself.  According to the Virial theorem (applied to the envelope alone rather than the whole star), the thermal energy, that reduces the value of the potential energy and makes the envelope closer to being unbound, is approximately one half the envelope's binding energy \cite{dewi00}, thus adding a factor $\frac{1}{2}$ to the left hand side of equation \ref{eqn:alphafull} and halving the derived values of $\alpha$ \cite{demarco11}.  However, the Virial theorem is based on the assumption of hydrostatic equilibrium, which may not apply if the CE happens on timescales much shorter than the stellar dynamical timescale.  As such, while it is important to account for this thermal energy when considering the binding energy of the envelope, one must exercise caution in including in it the calculation of the total energy and thus whether the envelope is ultimately unbound or not \cite{demarco11}.

As well as the envelope's thermal energy, its rotational energy is similarly neglected in the treatment outlined earlier.  This seems to be a reasonable assumption given that it should be negligible compared to its gravitational binding energy \cite{webbink08}.  A perhaps more questionable assumption is the treatment of the primary's core and the companion as inert masses, which do not gain or lose energy nor mass during the process.  It seems plausible to consider that the companion might accrete (appreciably) during the CE, however it has generally been thought that, due to the highly supersonic in-spiral and the large entropy barrier that forms between the secondary and the far less dense envelope, the total accretion is rather limited \cite{hjellming91,sandquist98}.  More recent models seem to indicate that this might not be the case providing there is some way of releasing pressure and thus maintaining steady flows during in-spiral \cite{ricker08,chamandy18}, the most obvious being the formation of jets which might also help to ``clear out'' the envelope increasing the ejection efficiency \cite{lopez-camara19,shiber19}.  Furthermore, if the companion is itself a compact object then nuclear burning  (perhaps of accreted material) on its surface could potentially provide a further energy source within the CE that might help to unbind the envelope \cite{iben93}.

As the CE is ejected preferentially in the orbital plane (due to the conservation of angular momentum), a pressure gradient develops which leads to material above and below the plane flowing inwards to replace that which has been ejected.  This circulation could plausibly lead to mixing that increases the primary's nuclear burning rate, providing additional energy which might contribute to the envelope's ejection \cite{iben93}.  However, it is seemingly unlikely that, in the face of strong entropy forces, the dynamical penetration of could reach deep enough to impact on the burning region \cite{webbink08}. 

The deposition of energy into the envelope (even just via gravitational drag) can have profound effects on its stability beyond simply lifting and ejecting it.  Modelling efforts have demonstrated that the envelope itself can rapidly become unstable and develop large-amplitude pulsations on relatively short timescales \cite{ivanova15,clayton17}.  In some cases, the shocks associated with these pulsations may be sufficient to dynamically eject up to 10\% of the primary's envelope \cite{clayton17}.

Perhaps the most contentious additional energy source in the CE is the inclusion of the envelope's recombination energy \cite{ivanova15,nandez16,grichener18,soker18}.  This energy source is significant, being proportional to the mass of the envelope and of order a few times 10$^{46}$ ergs c.f $\sim$10$^{47}$ for the binding energy of the envelope \cite{ivanova16,nandez16}.  However, it has been argued that only a small fraction of this energy ($\sim$10\%) might actually be able to contribute to the removal of the envelope, with the majority simply radiated away \cite{grichener18}. The true importance of recombination energy is still a matter of intense debate \cite{ivanova18,soker18}.  Recent studies of the impact of convection on the CE indicate that the associated energy transfer timescale is shorter than dynamical timescales, such that recombination energy may be convectively carried to the outer parts of the envelope where it is unable to aid with ejection \cite{wilson19}.  However, they also find that the inclusion of convection in their models could reduce the need for additional energy sources, as the it leads the binary orbit to shrink significantly before orbital energy can be tapped for ejection \cite{wilson19}.

Hydrodynamic models of the CE, that do not include recombination energy, invariably fail to unbind the entirety of the envelope, in the majority of cases leaving a significant fraction ``lifted'' away from the central binary but not completely unbound \cite{sandquist98,passy12a,ohlmann16,chamandy18,iaconi19,reichardt19}.  This led \cite{glanz18} to propose an intriguing solution, just as in single AGB stars \cite{lagadec08}, radiation pressure on grains which form in the now cooling (but still bound) envelope could lead to dust-driven winds capable of efficiently unbinding the envelope.  The authors also highlight that such dust formation could also aid in trapping any recombination energy from the envelope, adding weight to its possible importance in the CE ejection process.  While dust forming in the expanding common envelope likely has the ability to increase the mass-loss rate from the envelope and help with unbinding it, a question remains surrounding the timescales. Dust can form rapidly in the relatively high density common envelope gas. However, fall-back of bound material can happen equally rapidly \cite{kuruwita16}. The post-in-spiral environment would therefore see a complex competition of mechanisms including: the release of recombination energy, dust formation, fall-back of material. Somehow the competition of all these processes must result in the full envelope ejection, at least in some cases.

\subsection{The gamma prescription}
\label{sec:gamma}

An alternative to the $\alpha$ prescription has also been proposed based on the conservation of angular momentum \cite{nelemans00,nelemans05}.  Here, the same energy conservation as in the $\alpha$ formalism is implicit, although not restricted simply to orbital and binding energy.  This prescription can be written:
\begin{equation}
    \frac{\Delta J}{J} = \gamma \frac{\Delta M}{M} = \gamma \frac{M_e}{M_e+M_c+M_2}
\end{equation}
where $J$ and $\Delta J$ are the the total angular momentum of the binary and the change in total angular momentum, respectively.
 
This use of this formulation was driven by the apparent difficulty in explaining double WD binaries, which were thought to have formed from two consecutive CE episodes (though this has since been shown to not be necessary; see Sec.~\ref{sec:dd} \cite{woods12}).  Initial results were particularly encouraging, with reconstruction methods finding that a single value of $\gamma$ could well reproduce a number of systems \cite{nelemans00,nelemans05}.  The question then became, is this formulation intrinsically ``better'' than its $\alpha$ counterpart? And, moreover, could the apparently universal value of $\gamma$ be used to constrain the physics of the CE?  Unfortunately, it has been shown beyond doubt that the energy conserving prescription places far more constraints on the CE outcome \cite{webbink08}.  Ultimately, the limited range of $\gamma$ which can be used to successfully reconstruct all post-CE systems \cite{zorotovic10}, is rather more a short-coming of the formalism than a sign of insight into the CE physics.  This is highlighted by the small range of $\gamma$ required to account for essentially all possible post-CE configurations \cite{webbink08}.  This was shown mathematically by \cite{woods11} who demonstrated that the ratio of initial to final orbital separations is extremely sensitive to small changes in $\gamma$ -  with the range of $1.5\lesssim \gamma \lesssim 1.75$ found by \cite{nelemans00} encompassing values that would lead to merger during the first CE and values that could lead to a double WD system.  For further, in-depth discussion of failings of the $\gamma$ formalism as a replacement for $\alpha$, the reader is referred to Sect.~5.2 of \cite{ivanova13}.

\subsection{Grazing envelope evolution}
\label{sec:ge}

We have already mentioned the possibility that jets may help remove the envelope, as well as acting as a pressure-release valve and allowing for appreciable accretion during the CE \cite{chamandy18}. However jets could fundamentally alter the general picture of the CE as presented in figure \ref{fig:CE}.  As outlined by \cite{soker15}, one might envisage a situation where jets launched at the onset of the CE remove enough of the envelope to prevent engulfment and a full-blown CE event, rather maintaining the system in a constant state of ``just entering the CE phase''.  This hypothesis is known as the Grazing Envelope (GE).

The GE evolution could prevent a CE entirely or postpone it, removing a significant amount of the envelope prior to engulfment.  Clearly, this has a dramatic effect on the energetics of the CE as described above \cite{soker15}, principally providing a significant additional energy source proportional to the amount of mass accreted onto the secondary.  As such, the GE offers a, perhaps, more reasonable explanation for longer period post-CE binaries, such as the 16-d period binary central star of the PN NGC~2346 \cite{brown19} which \cite{soker15} argue must have experienced a GE for a significant fraction of its evolution.  Indeed, hydrodynamical simulations of the CE including jets do lead to greater final separations than than those without jets and also lead to a greater fraction of the envelope being unbound, however they still fail to completely unbind the entirety of the envelope \cite{shiber19}.

Beyond the additional energy source, the most significant difference between the CE and GE are the evolutionary timescales, with the GE expected to last tens to hundreds of years c.f. the days to months long duration of the standard CE.

\section{Planetary Nebulae nuclei}

As previously highlighted, the immediate product of the CE ejection is expected to be a PN with a close-binary nucleus\footnote{This may only be the case if the CE occurs on the AGB.  If the CE occurs while on the RGB, there are doubts as to whether the post-CE evolution of the exposed core would be fast enough to ionise the expanding envelope in time for it to be visible as a PN. However recent theoretical and observational efforts seem to indicate that these doubts are unfounded and that post-RGB PNe do indeed exist \cite{hall13,hillwig17}.}. As such, we may look to these objects for clues towards understanding the CE process.  It is only now, following the recent leaps in sample size (due to, for example, the OGLE survey \cite{miszalski09a} and recent targeted surveys \cite{jones14,jones15,santander-garcia15,jones19}), that one can begin to contemplate this idea \cite{jones17}.    In this section, I will try to discuss some of the more interesting findings to have come from the study of PNe in terms of the CE phase.

\subsection{Morphologies}
\label{sec:morph}

\begin{figure}[t]
\centering
\includegraphics[width=0.8\textwidth]{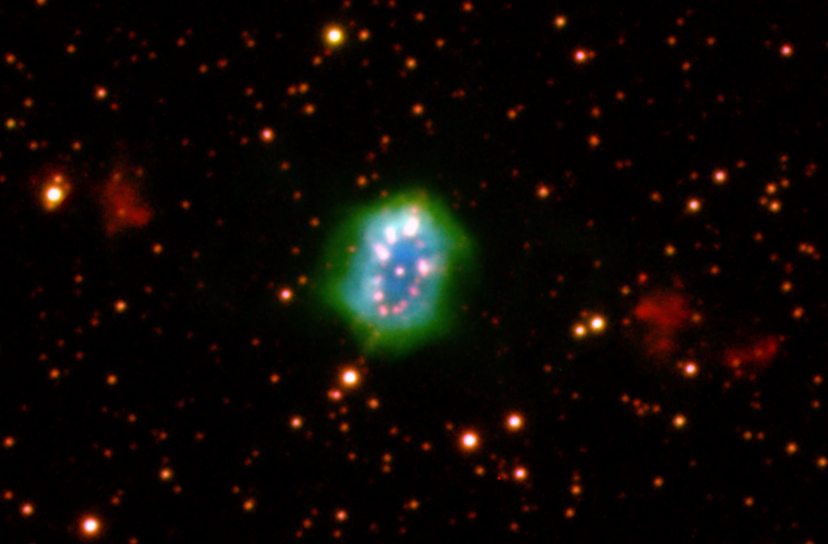}
\caption{Image of the Necklace Nebula (Credit: Romano L.~M. Corradi, IPHAS), shown to host a post-CE binary central star in which the companion is a carbon dwarf \cite{corradi11,miszalski13}. }
\label{fig:necklace}       % Give a unique label
\end{figure}

\begin{figure}[t]
\includegraphics[width=\textwidth]{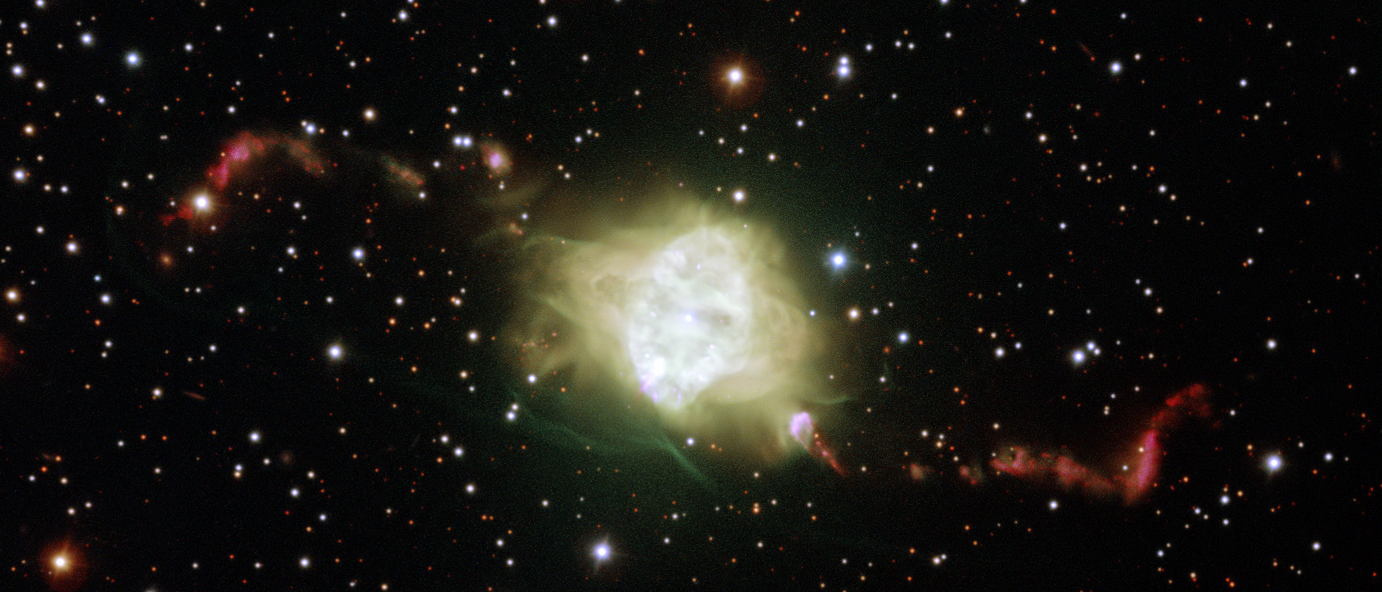}
\caption{FORS2 image of the post-CE PN Fg~1 (Credit: ESO/H. Boffin).  The observed jets have been shown to pre-date the central nebula by a few thousand years, while the central star is a double-degenerate binary with an orbital period of 1.195d \cite{boffin12}. }
\label{fig:fg1}       % Give a unique label
\end{figure}

The PNe surrounding post-CE central stars are thought to principally comprise the ejected envelope itself, thus they offer a unique window into the ejection process.  Studying the morphologies of post-CE PNe currently represents the \emph{only} way to observe the form of the CE when ejected in its entirety\footnote{The ejecta from luminous red novae, as discussed in Sec.~\ref{sec:lrn}, could also be used to study the CE ejection.  However, these objects represent ``failed'' CEs, where the binary merged inside the CE (likely ejecting only a fraction of the envelope).}.  There is an added complexity in inferring the structure of CE ejecta from post-CE PNe in that the observed morphologies are the result of a complex interplay between the initial ejecta and subsequent fast wind and ionisation front originating from the emerging pre-WD core \cite{garcia-segura18}.  As such it is perhaps no surprise that post-CE PNe display a wide range of morphologies, but with a few over-arching trends that can be used to understand the CE.  Most importantly, the vast majority of post-CE PNe present with bipolar morphologies \cite{miszalski09b}.  Spatio-kinematical modelling of these bipolar structures has revealed that their symmetry axes are always found to lie perpendicular to the orbital plane of the surviving post-CE central binary \cite{hillwig16}.  This is a clear confirmation that the CE is, indeed, preferentially ejected in the orbital plane, with the subsequent equatorial over-density going on the form the waist of the resulting bipolar nebulae - in many cases leading to the formation of a ring or torus (Figs.~\ref{fig:necklace} and \ref{fig:fg1} \cite{corradi11,boffin12}).

Post-CE PNe have also been found to show a prevalence of jet-like structures as well as low-ionisation filaments or knots. Jets are rather clearly a consequence of mass transfer which, given the relatively short nebular visibility times ($\tau\approx30,000$ years), must have occurred around the time of the CE (either just before, during or just after - a question we will return to in Sec.~\ref{sec:prece_masstransfer}).  The low-ionisation filaments and knots are slightly more challenging to understand, but might perhaps be related to instabilities in the envelope at the time of ejection (such as those described in \cite{ivanova15,clayton17}), or to a later fast tenuous wind originating from the central star ploughing into the ejected material.

\subsection{Double degenerates}
\label{sec:dd}

Unfortunately, while the number of PNe known to host post-CE central stars has grown dramatically, in the majority of cases little more is known beyond the orbital period.  Most systems were discovered via photometric monitoring, with the periodicity revealing the orbital period\footnote{With lower fidelity data it can be difficult to distinguish between variability due to irradiation and variability due to ellipsoidal modulation, plausibly leading to a derived orbital period which is discrepant by a factor of two \cite{manick15,boffin19}.} and the type of variability offering some hints towards the evolutionary phase of the secondary.  The majority of photometrically-discovered post-CE central stars display variability due to irradiation or ellipsoidal modulation, sometimes with eclipses superimposed if the orbital inclination is high enough.  All ellipsoidally-modulated central stars subjected to further detailed study have been found to be double-degenerate (DD) systems \cite{hillwig10,santander-garcia15}, where the companion is also an evolved star (WD or post-AGB star).  If one assumes that all the known post-CE central stars displaying ellipsoidal modulation in their light curves are indeed DDs, then they should comprise at least 20\% of the total population - likely far greater as such DD systems will only display photometric variability at very short orbital periods (indeed a number of further systems have been discovered via radial-velocity monitoring which do not display any photometric variability \cite{boffin12,miszalski18}).

Such a high fraction of DD central stars is particularly intriguing for a number of reasons.  As noted in Sec.~\ref{sec:gamma}, there is some debate over how such systems form - either via consecutive CE episodes or through stable mass transfer followed by a single CE, or perhaps even via a GE evolution.  The first possibility - consecutive CEs -  clearly presents a challenge, with the first CE already spiralling-in the binary and vastly reducing the orbital energy available to unbind a second CE (particularly problematic given that due to the reduced separation the second CE is more likely to occur with the donor on the RGB rather than AGB, where the envelope is more bound \cite{jones19imbase}).  It has been shown that under certain circumstances the first CE could be avoided, with the initially more massive component losing its envelope through a phase of stable, non-conservative mass transfer allowing the binary to continue towards being DD via a single CE episode \cite{woods12}.  However, it is unclear what range of initial conditions could lead to this evolution, and whether this range is sufficiently broad to account for the large DD fraction observed.  GE evolution can relatively easily explain such DD systems, however it is unclear what initial configurations (if any) could lead to some form of GE evolution.

A high DD fraction is particularly interesting in the context of understanding type \textsc{i}a supernovae (SNe) which - in spite of being successfully employed as standard candles in probing the increased expansion rate of the Universe \cite{riess98,perlmutter99} which ultimately led to the award of the Nobel prize in Physics 2011 - still have rather uncertain origins.  DD mergers may represent the main, or even sole, pathway by which SNe \textsc{i}a occur \cite{maoz14}, however to-date no bona-fide progenitor system has been discovered.  As such, a high DD fraction, as well as the observation that many SN \textsc{i}a are found to explode in circumstellar environments consistent with a remnant PN shell \cite{tsebrenko15}, could be construed as support for this DD merger hypothesis.  Furthermore, two of the strongest candidate SN \textsc{i}a progenitors have been found to reside inside PNe. However, neither has been unambiguously shown to satisfy the criteria of being both super-Chandrasekhar mass and in a close enough orbit to merge within the age of the Universe.  The central star of TS~01 was found to be a short-period DD but the total mass of the system is rather uncertain - encompassing both sub- and super-Chandrasekhar solutions \cite{tovmassian10}.  Somewhat similarly, simultaneous light- and radial-velocity curve modelling of the central star of Hen~2-428 led \cite{santander-garcia15} to conclude that the system was a DD with total mass 1.76$\pm$0.26 M$_\odot$ that would merge in roughly 700 million years.  While recent analyses have confirmed the DD classification \cite{finch18}, they have brought into question its super-Chandrasekhar nature with different spectral lines seemingly presenting with different radial velocity amplitudes \cite{reindl18}, and thus differing mass solutions (some of which are sub-Chandrasekhar).  In any case, both Hen~2-428 and TS~01 represent plausible SN \textsc{i}a progenitor candidates, and emphasise the possible importance that the high DD fraction among post-CE PNe may hold in evaluating the DD merger scenario for SNe \textsc{i}a in general.

\subsubsection{Mass and period distributions}
\label{sec:distributions}

Only a handful of post-CE central stars have been subjected to the detailed modelling required to derive their stellar parameters \cite{jones17,jones19,boffin19}.  However, those that have present with some intriguing properties - when one considers main sequence companions (see Tab.~\ref{tab:cb_modelled} for a list of post-CE central stars with MS secondaries and well-constrained masses, temperatures and radii), for example, all but one are found to present with rather low masses.  Indeed, other than in the case of Sp~1, the primaries are still more massive than the companions even after the ejection of the CE.  This is consistent with the studies highlighted in Sec.~\ref{sec:conditions} which indicate that, in order to experience dynamical Roche lobe overflow, the initial mass ratio ($q=M_1/M_2$) must be rather high.  A similar dearth of more massive companions is also found in the general post-CE population \cite{davis12}, with some suggesting that this may be due to the intrinsic difficulty in identifying white-dwarf-main-sequence (WDMS) binaries with a massive (and optically bright) MS component as a result of the large brightness difference between the two stars \cite{parsons16}.  This is not such an issue for PN central stars, where the pre-WD would be more luminous than a typical field WD which has already reached the cooling curve.  Furthermore, the presence of the nebula itself makes the identification of such systems more likely, acting as a signpost for the existence of a central white dwarf which would present with colours very different to that of earlier-type (AFG or even K type, for example) MS stars - thus such MS stars stand-out when found at the centre of a PN.  Several wide-binary central stars have been identified through this methodology, initially being labelled ``peculiar'' due to the discovery of the optically-bright companion before the identification of the nebular progenitor \cite{lutz77,aller15,jones17b}.  Only one such system has since been shown to be a post-CE binary - the central star of NGC~2346 - which is made even more unusual by playing host to a $\sim$3.5 M$_\odot$ sub-giant companion \cite{brown19}, making it the most massive post-CE secondary known (not just among post-CE PNe but in general).

\begin{table}
\caption{Close-binary WDMS central stars with well-constrained masses, radii and temperatures derived from simultaneous light and radial velocity curve modelling.}
\label{tab:cb_modelled}       % Give a unique label
\begin{tabular}{lccccccccl}
\hline\noalign{\smallskip}
PN & Period & M$_{CS}$ & R$_{CS}$ & T$_{CS}$ & M$_S$ & R$_S$ & T$_S$ & i  & Ref\\
& (days) & (M$_\odot$) & (R$_\odot$) & (kK)& (M$_\odot$) & (R$_\odot$) & (kK) &($^\circ$)&\\
\noalign{\smallskip}\svhline\noalign{\smallskip}
Abell~46 & 0.47 & 0.51$\pm$0.05 & 0.15$\pm$0.02 & 49.5$\pm$4.5 & 0.15$\pm$0.02 & 0.46$\pm$0.02 & 3.9$\pm$0.4  & 80.3$\pm$0.1 & \cite{afsar08}\\
Abell~63 & 0.47 & 0.63$\pm$0.05 & 0.35$\pm$0.01 & 78$\pm$3 & 0.29$\pm$0.03 & 0.56$\pm$0.02 & 6.1$\pm$0.2  &87.1$\pm$0.2 & \cite{afsar08}\\
Abell~65 & 1.00 & 0.56$\pm$0.04 & 0.056$\pm$0.008 & 110$\pm$10 &0.22$\pm$0.04 &0.41$\pm$0.05 & 5.0$\pm$1.0 & 61$\pm$5  & \cite{hillwig15}\\
DS~1 & 0.36 & 0.63$\pm$0.03 & 0.16$\pm$0.01 & 77$\pm$3 & 0.23$\pm$0.01 & 0.40$\pm$0.01 & 3.4$\pm$1 & 62.5$\pm$1.5 &  \cite{hilditch96}\\
ESO~330-9 &0.30&0.38--0.45&0.03--0.07&55--65&0.3--0.5&0.35--0.50&$\leq$4.5&7--13& \cite{hillwig17}\\
HaTr~7&0.32&0.50--0.56&0.13--0.18&90--100&0.14--0.20&0.3--0.4&$\leq$5&45--50& \cite{hillwig17}\\
Hen~2-155 & 0.15 & 0.62$\pm$0.05 & 0.31$\pm$0.02 & 90$\pm$5 & 0.13$\pm$0.02 & 0.30$\pm$0.03 & 3.5$\pm$0.5 & 68.8 $\pm$0.8 & \cite{jones15}\\
LTNF~1 & 2.29 & 0.70$\pm$0.07 & 0.08$\pm$0.01 & 105$\pm$5 & 0.36$\pm$0.07 & 0.72$\pm$0.05 & 5.8$\pm$0.3 & 84$\pm$1 & \cite{ferguson99}\\
M 3-1 &0.13 &0.65$^\star$ &0.41$\pm$0.02 & 48$^{+17}_{-10}$ & 0.17$\pm$0.02 & 0.23$\pm$0.02 & 5--12 & 75.5$\pm$2 & \cite{jones19}\\
NGC~6337 & 0.17 &0.56$^\star$ &0.045--0.085&115$\pm$5&0.14--0.35&0.30--0.42&4.5$\pm$0.5&17--23& \cite{hillwig16}\\
Sp~1 &2.91&0.52--0.60&0.20--0.35&80$\pm$10&0.52--0.90&1.05--1.60&3.5--4.6& 7--11 &  \cite{hillwig16}\\
\noalign{\smallskip}\hline\noalign{\smallskip}
\end{tabular}\\
$^\star$Fixed in the modelling\\
\end{table}

The distribution of primary masses is also interesting, with a number presenting masses consistent with being post-RGB objects \cite{hillwig17}.  This is perhaps in-keeping with the radius evolution properties of low-mass stars as discussed in Sec.~\ref{sec:conditions}. However, such stars were thought likely to evolve too slowly following the ejection of their envelopes and thus never produce a visible PN (only reaching the temperature required to ionise the envelope long after it has dissipated into the surrounding interstellar medium).  Recent theoretical studies of the end-of-CE structures of such stars have challenged this interpretation, indicating that core masses as low as 0.3~M$_\odot$ may well be capable of producing an observable PN \cite{hall13}.  As such, these post-RGB stars offer a promising avenue to understand the properties of such stars upon leaving the CE and thus the physics of RGB CEs.

In terms of the post-CE PN period distribution, it is generally very similar to that of the general WDMS population \cite{miszalski09a}, though perhaps with some minor differences which result from small number statistics and detection biases \cite{boffin19}.  In both cases, the period distribution shows a strong peak at around 8 hours, with a paucity of longer period systems (greater than a few days). Only IK Peg, FF Aqr, V1379 Aql and the central stars of NGC~2346 and MyCn~18 present with confirmed post-CE periods longer than 5 days\footnote{The double-degenerate central star of NGC~1360 has an orbital period of $\sim$142 days \cite{miszalski18}, but may not be the result of a CE.  Instead, such systems may evolve through stable, non-conservative mass transfer \cite{woods12,toonen13} as described in Sec.~\ref{sec:dd}.} \cite{davis10,brown19,miszalski18b}. IK Peg, FF Aqr, V1379 Aql and NGC~2346 also have the most massive secondaries known, perhaps hinting at a connection between secondary and/or primary mass and CE ejection efficiency (see \cite{iaconi19}).  Some attempts to reconstruct the CE phase for the general population of WDMS binaries with known masses and periods (roughly as described in Sec.~\ref{sec:alpha}) do indeed show statistically significant correlations between primary mass and ejection efficiency \cite{davis12}, and between the mass ratio and the efficiency \cite{demarco11}.  However, these are not universal findings, with other studies claiming that there are no dependencies \cite{politano07,zorotovic10} - see \cite{wilson19} for a critical overview of the possible correlations determined, and some reconciliation of the apparently contradictory results from previous studies.  In any case, population synthesis models seem to indicate that the efficiency must be low in order to explain the absence of longer period systems \cite{toonen13,camacho14}, which is very unlikely to be solely due to observational bias \cite{demarco08,nebot11}.

\subsection{Pre-common envelope mass transfer}
\label{sec:prece_masstransfer}

As highlighted in Sec.~\ref{sec:distributions}, only a handful of post-CE central stars with MS secondaries have been subjected to detailed modelling, but in spite of these they hold even more surprises beyond their mass distribution.  In all but one case, the MS companions were found to be greatly inflated (sometimes by a factor of two or more) with respect to isolated stars of similar masses \cite{jones15}, and even though the exception to this rule shows a fairly typical radius for its mass it was found to be Roche-lobe filling and thus could not be inflated without transferring material back onto the WD primary \cite{jones19}.  While some of this inflation could be due to the high levels of irradiation from the hot, nebular progenitors \cite{demarco08}, it is now generally accepted that it is principally a consequence of rapid mass transfer on to the MS star either during or just prior to the CE phase \cite{jones15}.  This mass transfer knocks the star out of thermal equilibrium causing them to ``puff up'' - a state in which they remain given that the thermal timescale of these stars is orders of magnitude longer than the timescale of the CE ejection as well as the PN visibility time.  In this way, the post-CE central stars of PNe are often referred to as ``fresh-out-of-the-oven'', as the relatively short PN lifetime (a few tens of thousands of years) guarantees that the central binary has had little time to adjust following the CE ejection.

Further support for significant accretion onto main sequence companions in post-CE central stars comes from the spectacular Necklace nebula (Fig.~\ref{fig:necklace}).  Discovered as part of the IPHAS survey \cite{drew05}, the central star was later shown to be a photometrically-variable post-CE binary with a period of 1.16d \cite{corradi11}.  Later spectroscopic observations taken around the photometric minimum revealed that the MS companion in the system was greatly enriched in carbon \cite{miszalski13}. Such carbon dwarfs are either primordially-enriched in carbon or are the product of chemical contamination via accretion from a more evolved companion -- the latter hypothesis being supported by the large number of dwarf carbon stars which are also found to be X-ray bright (considered a strong sign of post-accretion activity).  In any case, the carbon dwarf in the necklace is highly likely to have been contaminated via accretion given that nebula also presents with a remarkable pair of polar outflows or jets, which are almost certainly also a consequence of mass transfer between the binary components.

A significant fraction of post-CE PNe are found to display polar outflows similar to the Necklace \cite{miszalski09b}, the properties of which can be used to probe the mass transfer chronology and, for example, the magnetic fields associated with the accretion disc (assuming that magnetic fields are responsible for angular momentum transport and jet launching). Kinematical studies of the jets reveal that in almost all cases the jets pre-date the central nebular regions by a few thousand years\footnote{The ages referred to here are kinematical ages and, as such, represent the minimum ages for each component (i.e.\ the age assuming that the material was ejected ballistically and has not been slowed by interaction with the surrounding interstellar medium).}.  Given that the central regions are thought to represent the remnant of the ejected CE, this is strong evidence that the jets originate from a phase of pre-CE mass transfer.  This hypothesis is supported by apparent precession rate of the jets of Fg~1 (see Fig.~\ref{fig:fg1}).  The central star was shown to be a post-CE DD binary with an orbital period of 1.195d, while hydrodynamic models indicate that the precession rate of the bipolar rotating episodic jets is inconsistent with such a short period binary instead being associated with the pre-CE orbital period \cite{boffin12}.  The magnetic fields strengths (a few Gauss) and accretion rates (10$^{-5}$--10$^{-6}$ M$_\odot$ yr$^{-1}$) associated with the formation of these pre-CE jets are consistent with wind accretion shortly before the onset of Roche lobe overflow \cite{tocknell14}, providing important constraints on the pre-CE evolution.

It thus seems clear that most, if not all, post-CE central stars must have experienced some form of pre-CE mass transfer episode.  With this in mind, it is perhaps interesting to return some of the previously highlighted results -- particularly that the ``long-period'' post-CE systems (IK Peg, FF Aqr, V1379 Aql, NGC2346) highlighted in Sec.\ \ref{sec:distributions} would have had initial mass ratios much closer to unity than, for example, those short-period systems in Table \ref{tab:cb_modelled}.  We have already discussed in Sec.\ \ref{sec:conditions} that an extreme mass ratio is likely required for the RLOF to be unstable and thus for a CE to occur, however the indication that pre-CE mass transfer occurs in a majority of systems seems to hint that the RLOF may initially be at least somewhat stable (although non-conservative), perhaps via wind RLOF.  It is thus perhaps not unreasonable to surmise that the closer to unity the initial mass ratio, the longer this pseudo-stable, non-conservative pre-CE RLOF could be.  The longer this phase, the more mass could be lost in the form of jets or via the outer Lagrange points \cite{macleod18,reichardt19} or, at the very least, more mass that will be redistributed within the system.  It has already been suggested that such extended phases of pre-CE mass transfer could greatly impact on the in-spiral, leading to wider binary systems just as observed \cite{iaconi19}.

\subsection{Chemistry}
\label{sec:chem}

The chemical properties of post-CE PNe can also potentially be used to probe the CE phase.  Their chemical abundances trace the abundances in the envelope and, if measured with sufficient precision, could feasibly be used to probe the evolutionary phase of the donor upon entry into the CE.  Indeed, in a handful of cases, the abundance patterns of post-CE PNe have been shown to be consistent with the CE cutting short the AGB evolution of the nebular progenitor \cite{demarco08,jones14}, helping to constrain the pre-CE configuration.  

It has recently been shown that some post-CE PNe display highly anomalous abundances depending on the emission lines used to derive them.  For more than 70 years, it has been clear that the abundances of ionised nebulae differ depending on whether they are measured using recombination lines or the much brighter collisionally-excited lines \cite{wyse42} - becoming known as the ``abundance discrepancy problem''.  In the general PN population, abundances from recombination lines are found to be a factor of 2--3 greater than those from collisionally-excited lines \cite{wesson05}.  Post-CE PNe, however, tend to show even larger abundance discrepancy factors \cite{wesson18}, with some even reaching up to several hundred \cite{corradi15}.  Multiple explanations have been considered for the smaller discrepancies found in the general population of PNe and H\textsc{ii} regions, ranging from temperature variations to non-thermal electron energy distributions \cite{peimbert71,nicholls12}, however in the most extreme (post-CE) cases chemical inhomogeneities play a dominant role (for a more-detailed explanation of the derivation of chemical abundances in astrophysical nebulae, as well as the possible explanations for the abundance discrepancy, the reader is referred to the excellent review chapter by Jorge Garc\'ia-Rojas in this volume).  

The chemical inhomogeneities in short-period post-CE PNe manifest themselves as a second, lower-temperature, higher-metallicity gas phase in addition to a more standard temperature and metallicity phase consistent with that of the general PN population.  In a majority of cases, this higher-metallicity gas is found to be centrally concentrated and closer to the central star \cite{garcia-rojas16,wesson18}.  This has led some authors to consider that it may represent a form of post-CE eruptive event which leads to the ejection of reprocessed material \cite{jones16} - a particularly intriguing prospect given that the overall abundance pattern of the higher metallicity ejecta is reminiscent of that of Neon novae \cite{wesson08}.  If this were the case then, presumably, the two gas phases would show differing kinematical properties.  Unfortunately, studying the kinematics of the second gas phase is challenging due to the intrinsic faintness of the recombination lines in which it is brightest, however a preliminary study comparing lines across different chemical species did find evidence for discrepant kinematics between the two gas phases \cite{richer17}.  Returning to the possibility that the high-metallicity gas originates from some form of reprocessing event, it is unclear what could lead to this eruptive event. However, it could speculatively occur as a result of fall-back of CE material \cite{kuruwita16}.  What does seem to be clear is that whatever process leads to the ejection of this higher-metallicity material, it only occurs in the very shortest-period post-CE binaries \cite{wesson18}, with longer-period post-CE PNe ($P_\mathrm{orb}\gtrsim$1.2 days) tending to present with less extreme abundance discrepancies.

\section{Mergers}
\label{sec:lrn}

\begin{figure}[t]
\sidecaption[t]
\includegraphics[width=0.63\textwidth]{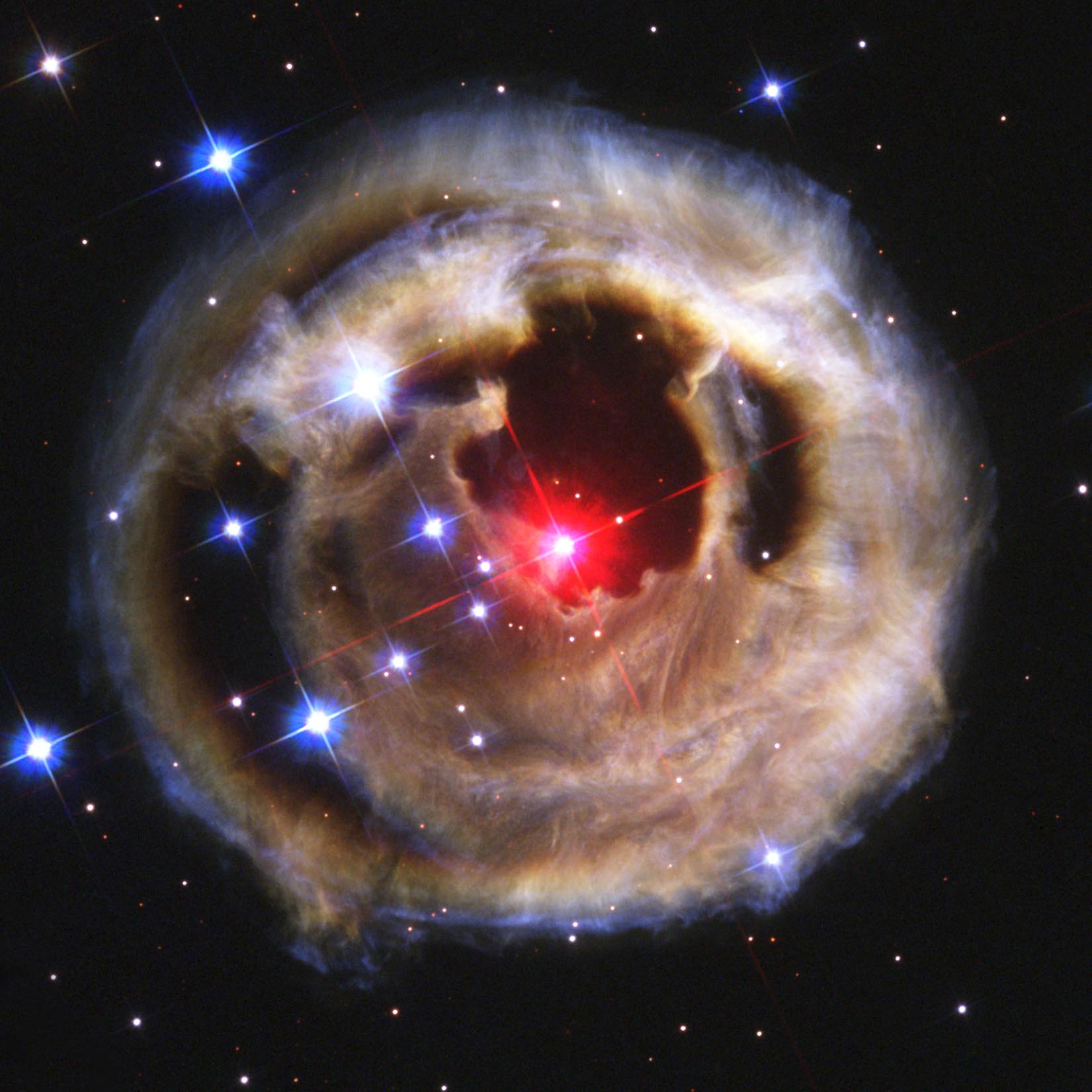}
\caption{Hubble Space Telescope image of the light echo around V838 Mon taken in October 2002 some nine months after eruption \cite{bond03}.  At this stage in its evolution, the central merger remnant presented with a temperature and radius consistent with an L-type supergiant \cite{evans03}. Image credit: NASA, ESA and H.E.~Bond (STScI). }
\label{fig:v838mon}       % Give a unique label
\end{figure}

Thus far, we have focused on systems which survived the CE phase as binaries, but a significant number will instead lead to mergers \cite{kochanek14}.  To date, only one PN central star can be considered a strong post-merger candidate - that of NGC~6826, found to display a rotation rate too high to have originated from a single star \cite{handler13,demarco15}.  However, another class of post-merger phenomena exist - luminous red novae\footnote{Continuing with the unfortunate misnomers surrounding CE-related phenomena, just as planetary nebulae have no relation to planets, lumninous red novae are completely unrelated to classical novae or supernovae.} (LRNe) - slowly-evolving red transients, the peak brightness of which is brighter than classical novae but fainter than supernovae (10$^{39}$--10$^{41}$ ergs s$^{-1}$ \cite{kasliwal12}).  At the turn of the century, only a handful of such transients had been identified and to relatively little fanfare - one (M~31-RV) being classified simply as ``a nova of unusual type'' which did not ``comfortably fit into the standard scenarios for eruptive events on white dwarfs'' \cite{mould90}.  However, our understanding of these transients was greatly advanced following the 2002 eruption of V838 Monocerotis (Fig.~\ref{fig:v838mon}) which, being detected early and residing in our own Galaxy, could be studied in exquisite detail \cite{bond03,evans03,munari05,munari07,chesneau14}.  The observed evolution of V838 Mon (as well as the other members of the class M~31-RV and V4332 Sgr) - a  brightening of several magnitudes followed by a slow decline, all the while developing redder and redder colours (V838 Mon resembling an L-type supergiant less than a year after its discovery \cite{evans03}) - was found to only be consistent with a merger scenario \cite{tylenda06,soker06}.  Neither a nova-like event (comprising some form of thermonuclear runaway on the surface of a WD) nor a helium shell flash or very late thermal pulse associated with a born-again event could be reconciled with the observed post-eruption colour/temperature evolution \cite{tylenda06}.

\begin{figure}[t]
\includegraphics[width=\textwidth]{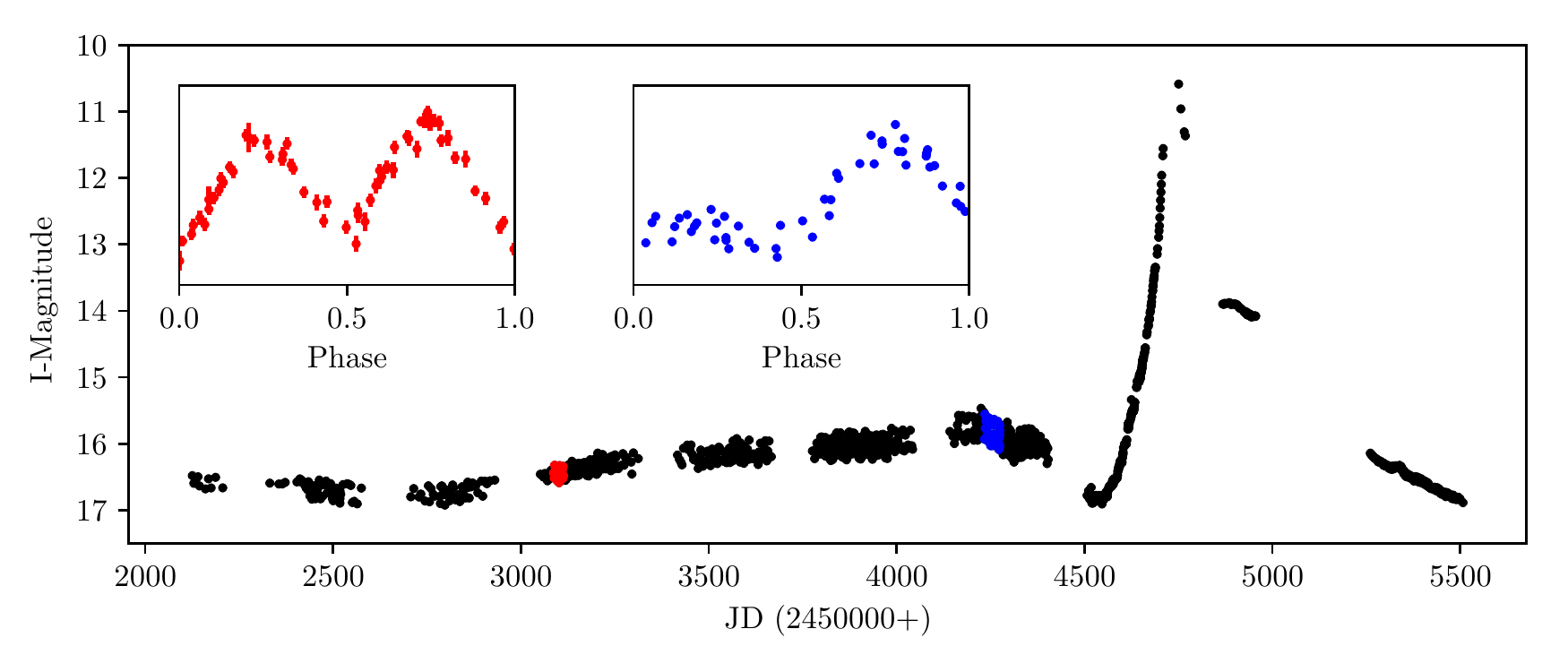}
\caption{The OGLE light curve of V1309 Sco showing the slow rise before eruption (data originally presented in \cite{tylenda11}.  Note that the final outburst was even brighter than shown here, reaching lower than 8$^{th}$ magnitude.  The insets show phase folded light curves corresponding to the regions of the same colour in the main plot, highlighting the evolution from double-peaked (the typical ellipsoidal modulation of a Roche-lobe filling binary) through to single-peaked (where the previous secondary peak has now been obscured by the mass lost through the L$_2$ point \cite{pejcha17}).}
\label{fig:v1309sco}       % Give a unique label
\end{figure}

Following the LRN eruption of V1309 Sco, the merger scenario for these objects was confirmed with the pre-eruption light curve from OGLE showing clear evidence that the progenitor was a contact binary (red inset of Fig.~\ref{fig:v1309sco}), the orbital period of which was exponentially decaying \cite{tylenda11}.  These pre-merger observations offered a unique window into the processes that led up to the dynamical CE event in this system.  In the two years prior to eruption, the light curve evolved from showing the typical ellipsoidal modulations of a contact binary through to showing only a single peak.  This is likely due to obscuration of the binary by mass lost from the second Lagrange point \cite{pejcha17}.  Furthermore, roughly 200 days prior to the outburst, variability associated with the orbital period was no longer detected, instead being replaced by a systematic slow brightening, which again could be associated with mass loss through the outer Lagrange point during the final stages of orbital decay \cite{pejcha14}.  The subsequent rapid brightening ($\sim$4 mag in $\sim$5 days) was then likely due to the final, dynamical merger of the two components and the liberation of recombination energy associated with expulsion of a shell of CE material \cite{ivanova13b}.  Intriguingly, the total mass loss associated with the gradual in-spiral phase (several hundredths of a solar mass over the course of a few thousand orbits) is comparable to the amount thought to be ejected during the final merger phase \cite{pejcha17}.  This is a clear demonstration that pre-CE interactions may play an important role in the outcome of the CE itself - helping to remove the envelope or at least dramatically altering the initial conditions (mass ratio, envelope structure, etc.) prior to the dynamical interaction as compared to those typically employed in hydrodynamical simulations (see e.g.\ \cite{reichardt19}).  Furthermore, magnetic fields could be generated or amplified via shearing motions as a result of loss of corotation during the pre-CE phase of period decay \cite{regos95,pejcha17}.  These magnetic fields could feasibly launch jets removing further mass from the system prior to the dynamic event \cite{tocknell14}.

Beyond pre-merger observations such as those serendipitously obtained for V1309 Sco, their post-CE light curves can also provide important constraints on the CE process. Even a relatively simple model, whereby the emission from a spherically-symmetric ejection is controlled by a recombination front as the material cools, was shown to match the observed colour and luminosity evolution of the handful of LRNe known at the time \cite{ivanova13b}.  More realistic models of this process -- combining both three-dimensional magneto-hydrodynamics and radiation transport -- have the potential to reveal much more about the CE process \cite{galaviz17,iaconi19b,zhu19}.  However, such models are computationally particularly challenging, not only due to the difficulty in incorporating radiation transport into the chosen hydrodynamic modelling code but also due to the rather long wall clock times that are required \cite{galaviz17}.  In spite of this, important progress has been made, strongly indicating that continued effort may prove key in understanding the long-term post-ejection behaviour of LRNe \cite{iaconi19b}.

Finally, late-time observations of LRNe can be used to directly measure the amount of mass ejected during the merger, as well as its morphology and kinematics \cite{kaminski18}.  Similarly, such late-time observations can also be used to probe the nucleosynthesis which occurs during merger via the astrochemistry of the ejected envelope \cite{kaminski17}, placing constraints on how mergers could impact the enrichment of the interstellar medium.

\section{Discussion}
\label{sec:discussion}

In this chapter, we have introduced some of the fundamental concepts important for our understanding of the CE -- one of the most critical, yet poorly understood phases of close-binary evolution.  Subsequently, we have discussed some of the recent observational findings related to two CE phenomena -- post-CE PNe and LRNe.  Both show strong evidence for (appreciable) mass transfer/loss before the dynamical CE event, which may prove to be a critical ingredient in deriving self-consistent and complete models of the CE process \cite{macleod18}, particularly given that most hydrodynamical modelling efforts focus on the dynamical in-spiral and begin with the companion at the surface of the giant \cite{passy12a,ohlmann16}.

Post-CE PNe, which represent the direct progeny of CE events where the envelope was successfully ejected without the cores merging, also have more to tell us about the CE phase.  The spatio-chemical properties of the nebulae offer some indication that the final stages of CE ejection, particularly in systems with small final orbital separations, may result in some form of reprocessing and the ejection of chemically-enriched material into the expanding envelope (giving rise to the extreme abundance discrepancies observed in these systems).  This similarly indicates that modelling efforts need, not only to begin before the dynamical in-spiral phase but also, to extend out towards the nebular phase \cite{reichardt19}.  Whatever the process behind the ejection of this enriched material, it could feasibly play a role in unbinding the remaining envelope and successfully terminating the CE -- a problem faced by a majority of hydrodynamic models which generally fail to unbind the entirety of the envelope, instead leaving a fraction ``lifted'' but still bound to the central binary \cite{iaconi17}.

The mass distribution of the companions inside post-CE PNe (as well as the general post-CE WDMS population) seems to indicate that only systems with much more massive donors will experience, or at least survive, a CE - helping constrain the conditions for dynamically unstable Roche lobe overflow.  Understanding the initial parameter space which could lead to a CE is of critical importance, along with constraining the efficiency (or refining whichever prescription is chosen to derive the end result of a CE), for the population synthesis efforts which will prove essential in interpreting the awaiting deluge of close-binary phenomena that will be revealed by the Large Synoptic Survey Telescope (LSST \cite{LSST}) and the Laser Interferometer Space Antenna (LISA \cite{shah14,kalomeni16,toloza19}).

While we are still far from understanding the CE, the (thus far) limited observational studies of CE phenomena like post-CE PNe and LRNe have already provided valuable insight.  Continued, deeper study of these phenomena will, without doubt, further refine our understanding of the CE and, when combined with continued theoretical and computational modelling efforts, perhaps lead to a unified picture of the processes at work in the phase.

\begin{acknowledgement}
DJ would like to thank the referee, Orsola De Marco, for her comprehensive report which improved both the clarity and completeness of this review.

DJ acknowledges support from the State Research Agency (AEI) of the Spanish Ministry of Science, Innovation and Universities (MCIU) and the European Regional Development Fund (FEDER) under grant AYA2017-83383-P.  DJ also acknowledges support under grant P/308614 financed by funds transferred from the Spanish Ministry of Science, Innovation and Universities, charged to the General State Budgets and with funds transferred from the General Budgets of the Autonomous Community of the Canary Islands by the Ministry of Economy, Industry, Trade and Knowledge.
\end{acknowledgement}

\bibliographystyle{spphys}
\bibliography{references.bib}

%http://www.issn.org/en/node/344

\end{document}